\documentclass{ieeetj}
\usepackage{cite}
\usepackage{amsmath,amssymb,amsfonts}
\usepackage{algorithmic}
\usepackage{graphicx,color}
\usepackage{textcomp}
\usepackage{xcolor}
\usepackage{hyperref}
\hypersetup{hidelinks}
\usepackage{algorithm,algorithmic}

\usepackage{multirow}

\newcommand{\red}[1]{\textcolor{red}{#1}}

\def\BibTeX{{\rm B\kern-.05em{\sc i\kern-.025em b}\kern-.08em
    T\kern-.1667em\lower.7ex\hbox{E}\kern-.125emX}}
\AtBeginDocument{\definecolor{tmlcncolor}{cmyk}{0.93,0.59,0.15,0.02}\definecolor{NavyBlue}{RGB}{0,86,125}}

\def\OJlogo{\vspace{-4pt}$<$Society logo(s) and publication title will appear here.$>$}
\def\seclogo{\vspace{10pt}$<$Society logo(s) and publication title will appear here.$>$}

\def\authorrefmark#1{\ensuremath{^{\textbf{#1}}}}

\begin{document}
\receiveddate{XX Month, XXXX}
\reviseddate{XX Month, XXXX}
\accepteddate{XX Month, XXXX}
\publisheddate{XX Month, XXXX}
\currentdate{XX Month, XXXX}
\doiinfo{XXXX.2022.1234567}

\markboth{}{Author {et al.}}

\title{Test-time Cost-and-Quality Controllable Arbitrary-Scale Super-Resolution\\ with Variable Fourier Components}

\author{Kazutoshi Akita\authorrefmark{1}, Norimichi Ukita\authorrefmark{1}}
\affil{Toyota Technological Institute, Japan}
\corresp{Corresponding author: Kazutoshi Akita (email: sd21501@toyota-ti.ac.jp).}

\authornote{This work was partly supported by JSPS KAKENHI Grant Numbers 19K12129 and 22H03618.}


\begin{abstract}
Super-resolution (SR) with arbitrary scale factor and cost-and-quality controllability at test time is essential for various applications.
While several arbitrary-scale SR methods have been proposed, these methods require us to modify the model structure and retrain it to control the computational cost and SR quality.
To address this limitation, we propose a novel SR method using a Recurrent Neural Network (RNN) with the Fourier representation.
In our method, the RNN sequentially estimates Fourier components, each consisting of frequency and amplitude, and aggregates these components to reconstruct an SR image.
Since the RNN can adjust the number of recurrences at test time, we can control the computational cost and SR quality in a single model: fewer recurrences (i.e., fewer Fourier components) lead to lower cost but lower quality, while more recurrences (i.e., more Fourier components) lead to better quality but more cost.
Experimental results prove that more Fourier components improve the PSNR score. Furthermore, even with fewer Fourier components, our method achieves a lower PSNR drop than other state-of-the-art arbitrary-scale SR methods.
\end{abstract}




\maketitle


\section{INTRODUCTION}

\begin{figure}[t]
    \centering
    \includegraphics[width=1.0\columnwidth]{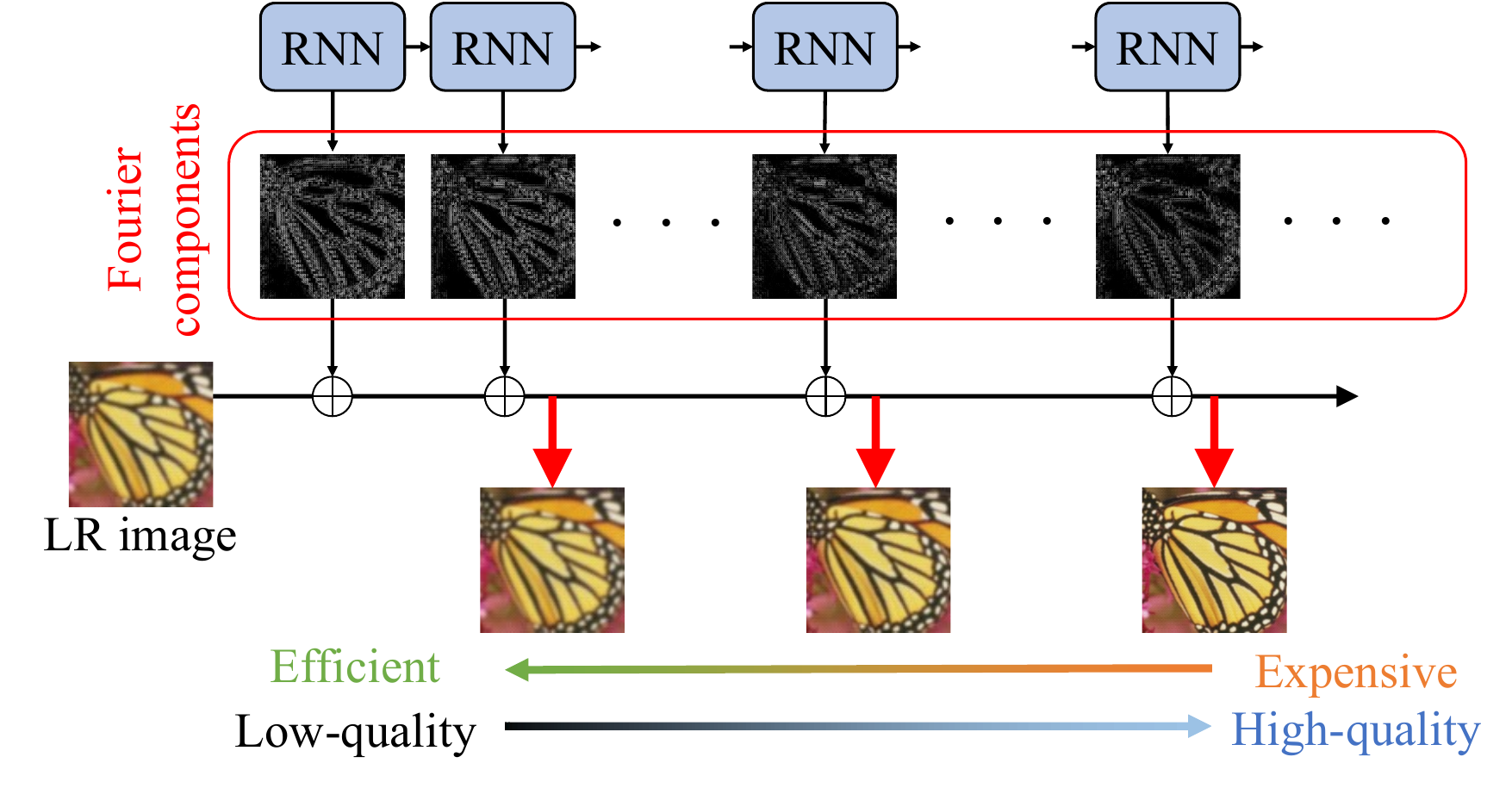}
    \caption{Conceptual illustration of the proposed CQ controllable arbitrary-scale SR. The RNN-based network estimates Fourier components one by one. By adjusting the number of RNN recurrences, the model can control the trade-off between computational cost and SR quality.}
    \label{fig:concept}
\end{figure}

\IEEEPARstart{S}{uper-resolution} (SR), the task of reconstructing High-Resolution (HR) images from Low-Resolution (LR) counterparts, is important for many applications, such as medical imaging, security systems, and satellite imaging. Early progress in SR~\cite{edsr},\cite{dbpn},\cite{wdst},\cite{srflow},\cite{pams},\cite{carn},\cite{lattice},\cite{srntt},\cite{spsr} came from deep learning, especially through Convolutional Neural Networks (CNNs). However, CNN-based methods are limited to integer scale factors, which limits their practical use.

This limitation is eliminated by recently proposed arbitrary-scale SR methods~\cite{metasr},\cite{Wang21},\cite{Vas23},\cite{Wang23arb},\cite{liif},\cite{lte},\cite{thera},\cite{ciaosr},\cite{clit},\cite{ultrasr},\cite{ipe}. However, for SR methods widely applicable to various real-world scenarios, not only arbitrary scale factors but also Cost-and-Quality (CQ) control are vital. CQ control in SR is the ability of an SR model to control its computational cost and SR quality depending on the given condition, such as the computational resource and the user requirement. CQ control is important in applications where the computational requirements vary, such as video streaming services and surveillance camera systems.

For CQ controllability, however, previous SR methods~\cite{liif},\cite{lte},\cite{ciaosr},\cite{clit} require us to adjust the model's complexity (such as the depth of network layers) and retrain the adjusted model. This requirement makes it impossible to control CQ with a trained SR model given to end users.

This paper proposes a novel SR method that simultaneously achieves arbitrary scale factors and CQ control without any modification in the SR model structure and additional training.
Our method achieves test-time CQ control in arbitrary-scale SR by utilizing a Recurrent Neural Network (RNN) to sequentially estimate an variable number of Fourier components (i.e., a pair of frequency and amplitude) used for SR image reconstruction, as shown in Fig.~\ref{fig:concept}.
The Fourier representation has many useful properties and is used in various low-level vision tasks~\cite{siren},\cite{Wu23},\cite{bacon},\cite{lte}. For example, image compression~\cite{compression1},\cite{compression2} by removing human-negligible components and image enhancement~\cite{Mao23},\cite{Wang23FaceSR},\cite{Fuoli21} by explicitly enhancing high-frequency components. Among them, we focus on the additive nature of Fourier components.
With this additive nature, the signal can be reconstructed from the sum of a variable number of Fourier components, while its reconstruction quality differs depending on the number of Fourier components used.

Our method adopts an RNN, which can handle variable-length outputs, to estimate a variable number of Fourier components one by one.
The estimated Fourier components are added up to reconstruct a final SR image. 
The more Fourier components, the more accurate the reconstructed SR image is.
On the other hand, the computational cost can be reduced by decreasing Fourier components.
With this property, our proposed method can achieve CQ controllability in a single SR model.

Our contributions are summarized as follows:
\begin{itemize}
    \item This paper proposes a test-time CQ controllable arbitrary-scale SR method.
    \item Our method achieves CQ controllable SR by using an RNN to estimate Fourier components, which are then used to reconstruct SR images. Unlike previous arbitrary-scale SR methods that fix CQ, our method can estimate and use a variable number of Fourier components to control CQ at test time.
    \item Our proposed training strategy accepts a variable number of Fourier components to allow our model to be generalized well across different numbers of Fourier components.
    \item Our experiments comprehensively evaluate the CQ controllability of our method compared to state-of-the-art arbitrary-scale SR methods. These results also demonstrate the effectiveness of maintaining high SR quality even with a varying number of Fourier components.
\end{itemize}


\section{RELATED WORK}

\subsection{Super-Resolution}

SR has been significantly advanced with deep neural networks. Early approaches, such as EDSR~\cite{edsr}, DBPN~\cite{dbpn}, WDST~\cite{wdst}, SRFlow~\cite{srflow}, PAMS~\cite{pams}, CARN~\cite{carn}, LatticeNet~\cite{lattice}, SRNTT~\cite{srntt}, and SPSR~\cite{spsr}, leverage CNNs. However, these methods are constrained by integer scale factors, such as 2, 3, and 4.
That is, SR with arbitrary scale factors, such as 2.1 and 5.23, is impossible.
This constraint limits their practical applicability.

This limitation is addressed by arbitrary-scale SR. MetaSR~\cite{metasr}, the pioneering approach in this domain, dynamically predicts upscale filters to handle arbitrary scale factors. 
Similarly, upsampling filters are dynamically predicted in several arbitrary-scale SR methods~\cite{Wang21},\cite{Vas23},\cite{Wang23arb}.
Alternatively, arbitrary scale SR is achieved by blending the SR images of integer scale factors, as proposed in~\cite{Son21},\cite{Wang19fusion}. 

The most recently-prevalent arbitrary-scale SR approach is based on Local Implicit Image Function (LIIF~\cite{liif}). This approach constructs an implicit function that estimates the RGB value of each pixel from its pixel position and its nearest latent code obtained from an input LR image. LTE~\cite{lte} and Thera~\cite{thera} predict Fourier components by the LIIF-based framework. CiaoSR~\cite{ciaosr} and CLIT~\cite{clit} flexibly aggregate latent codes beyond the nearest one. UltraSR~\cite{ultrasr} and IPE~\cite{ipe} replace the input pixel position of LIIF with the embedded one.

Despite the aforementioned innovative approaches, these methods require modifications and retraining to the SR model for CQ control.

\subsection{Fourier Representation}

The Fourier representation has been widely used in many low-level vision tasks. One primary application of the Fourier representation is to avoid spectral bias~\cite{spectral}, a phenomenon in which a neural network tends to be biased to learn low-frequency components. To address the spectral bias, SIREN~\cite{siren} utilizes periodic functions (e.g., sinusoidal function) as the activation function. Wu et al.~\cite{Wu23} use wavelet-like multi-scale components as the Fourier representation. BACON~\cite{bacon} sequentially predicts the Fourier components from low-frequency to high-frequency components.
LTE~\cite{lte} firstly adopts the Fourier representation in implicit functions for SR.

Another significant use of the Fourier representation is to extract rich image features through the Fourier Transforms. Mao et al.~\cite{Mao23} use the Fourier Transforms to extract global context features efficiently. Wang et al.~\cite{Wang23FaceSR} learn the interrelationship between spatial and frequency domains for incorporating local and global features. Fuoli et al.~\cite{Fuoli21} take reconstruction losses in the Fourier space to enhance the high-frequency features.

While these methods demonstrate the effectiveness of the Fourier representation in addressing the key challenges in low-level vision tasks, our proposed method is the first to leverage the Fourier representation to achieve test-time CQ controllable SR.

\subsection{CQ Controllable Neural Networks}

CQ control in neural networks is possible through various approaches. One prominent approach is knowledge distillation~\cite{distillation1},\cite{distillation2}, where a large pre-trained network (i.e., teacher model) transfers its knowledge to a smaller network (i.e., student model) that is designed for a given condition, such as the computational resource and the user requirements. However, knowledge distillation requires re-training each student network, which may be impractical for widespread deployment.

Other approach for CQ control include pruning~\cite{pruning1},\cite{pruning2} and quantization~\cite{quantize1},\cite{quantize2}.
After a large network is trained, these pruning and quantization approaches reduce its capacity by pruning less important connections and quantizing weights to lower precision (e.g., from 32 bits to 8 bits), respectively. That is, these pruning and quantization procedures are required for each given condition after the base large network is trained.
Furthermore, to maintain the performance even with a significantly reduced network capacity, data-driven pruning methods~\cite{data_driven1},\cite{data_driven2},\cite{data_driven3} and extra training on the reduced network are additionally required~\cite{pruningSV},\cite{quantizeSV}. Therefore, pruning and quantization are also impractical for widespread deployment.

While the aforementioned approaches require retraining, the early exit approach enables test-time CQ control without retraining. In the early exit approach, a large network is trained with output layers at intermediate layers. These output layers take features extracted through different numbers of layers.
For test-time CQ control, the network can terminate in any of these intermediate layers at test time. Huang et al.~\cite{huang18} make the Densenet capable of the early exit approach. Larsson et al.~\cite{larsson17} change the network depth using a fractal network design. Hu et al.~\cite{hu19} demonstrate that the weight of a loss function in each output layer is important in the early exit approach. Lin et al.~\cite{LinLHQ0W23} control CQ in SR by switching data flow in the residual model.

While our proposed method is similar to the early exit approach, our method focuses on the additive nature of Fourier components for CQ control.
The early exit approach uses independently trained intermediate output layers for CQ control. 
Due to this independent training, these layers may predict inconsistent results even across neighboring layers.
In contrast, our proposed method can produce predictions that are consistent according to variation in CQ. This is because recurrent predictions obtained until the $(t-1)$-th recurrence are used together with a prediction at the $t$-th recurrence to reconstruct the SR image at the $t$-th recurrence. That is, since all recurrently predicted Fourier components are just added up with no independent training, consistency is guaranteed between the SR images reconstructed at $(t-1)$-th and $t$-th recurrences.
Furthermore, the RNN is more efficient (i.e., with fewer parameters) than deep-layered networks for the early exit approach. Such a lightweight network is preferable for widespread deployment.


\section{METHODS}
\label{sec:methods}

\begin{figure}[t]
    \centering
    \includegraphics[width=1.0\columnwidth]{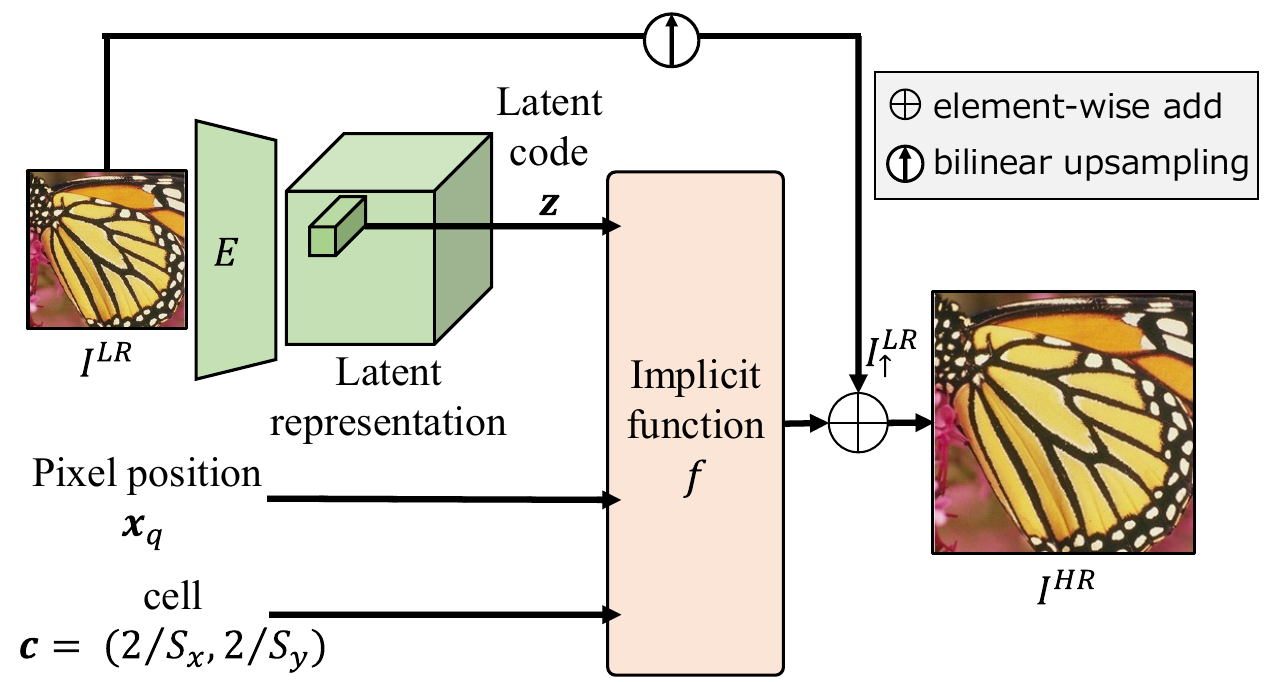}
    \caption{Overview of LIIF-based arbitrary-scale SR methods. The implicit function, $f$, is designed for each method.}
    \label{fig:overview}
\end{figure}

\begin{figure}[t]
    \centering
    \includegraphics[width=0.6\columnwidth]{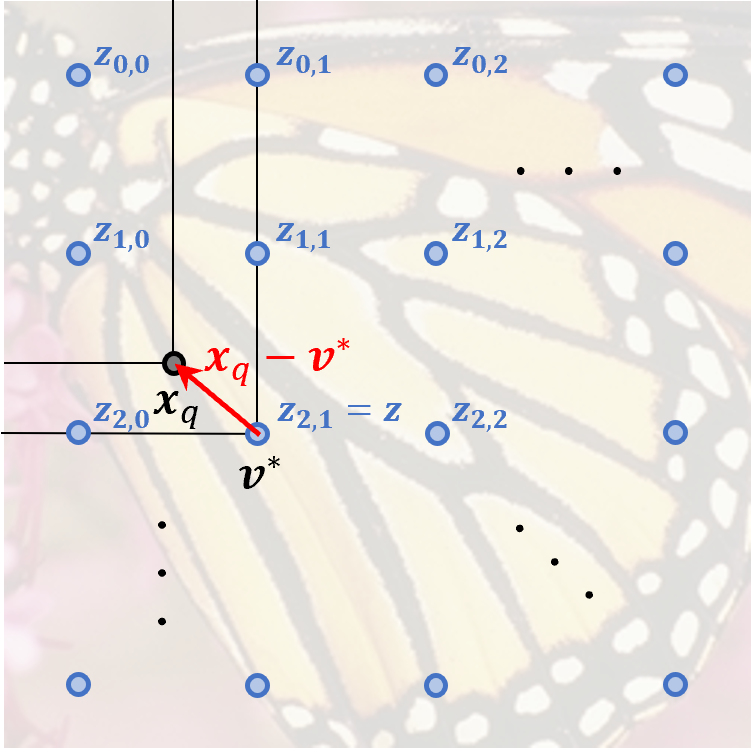}
    \caption{Coordinate system in LIIF-based methods. The blue-colored symbols indicate the latent code, while black-colored symbols indicate the position. $\mathbf{z}_{i,j}$ denotes the latent code at $(i, j)$, where $i$ and $j$ are the indices of the latent representation. $\mathbf{z}_{i,j}$ is tiled in the HR coordinate system at regular intervals. The latent code nearest to the query position, $x_q$, is selected as $\mathbf{z}$, and its position in the HR coordinate system is denoted as $\mathbf{v}^*$.}
    \label{fig:coord_system}
\end{figure}

\begin{figure}[t]
    \centering
    \includegraphics[width=1.0\columnwidth]{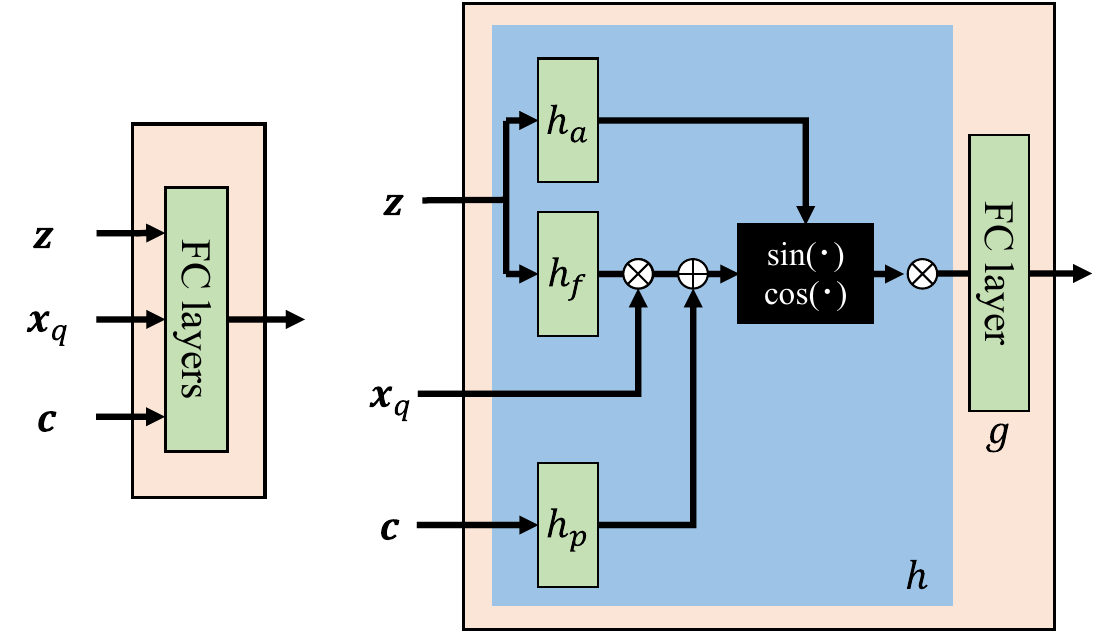}

    (a) LIIF~\cite{liif} \hspace*{25mm} (b) LTE~\cite{lte} ~~~~
    \caption{Implementations of the implicit function, $f$, in (a) LIIF~\cite{liif} and (b) LTE~\cite{lte}.}
    \label{fig:existings}
\end{figure}

\begin{figure*}[t]
    \centering
    \includegraphics[width=0.8\linewidth]{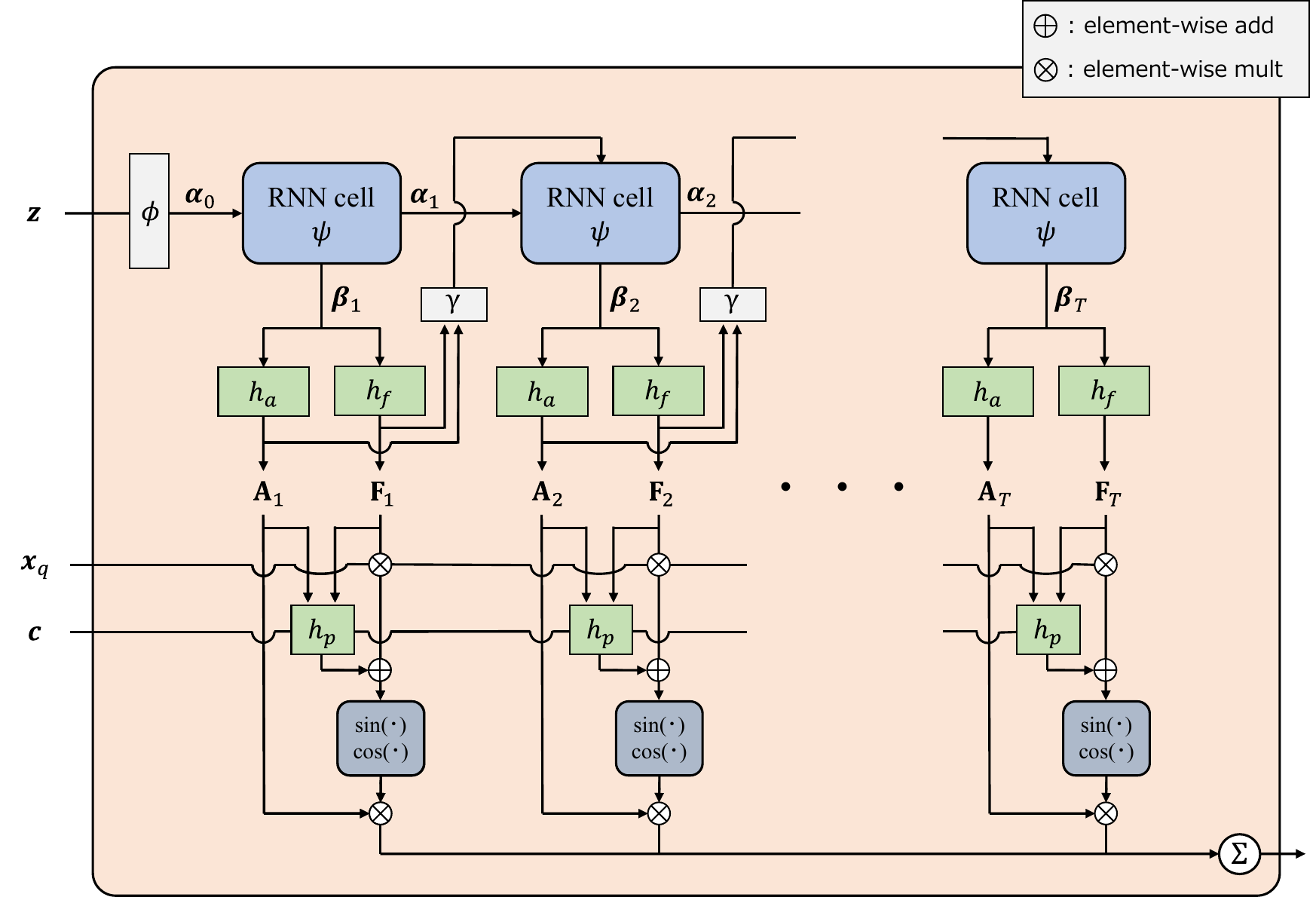}
    \caption{Our proposed implicit function. $\boldsymbol{\alpha}_t$ and $\boldsymbol{\beta}_t$ denotes the hidden state and output of RNN at the $t$-th recurrence. The initial hidden state $\boldsymbol{\alpha}_0$ is obtained from the latent code $\mathbf{z}$ through the function $\phi$. $\boldsymbol{\beta}_t$ is fed into the amplitude estimator $h_a$ and frequency estimator $h_f$, and obtain $\mathbf{A}_t$ and $\mathbf{F}_t$. After that, $\mathbf{A}_t$ and $\mathbf{F}_t$ are transformed by $\gamma$ and then passed to the next RNN recurrence. Additionally, phase information is estimated from the cell size, $\mathbf{c}$, $\mathbf{A}_t$, and $\mathbf{F}_t$, as described in Sec.~\ref{sec:methods}-\ref{sec:others}.
    Since the number of RNN recurrences $T$ can be adjusted freely at the test time, our proposed method can produce a variable number of Fourier components. These estimated Fourier components are summed to produce the final RGB value.}
    \label{fig:proposed}
\end{figure*}

\subsection{Preliminary and Motivation}

In SR, the goal is to obtain an HR image, $I^{HR} \in \mathbb{R}^{S_{y}H \times S_{x}W \times 3}$, from its LR image, $I^{LR} \in  \mathbb{R}^{H \times W \times 3}$, with a given scale factor, $S_y$ and $S_x$. Recent arbitrary-scale SR methods~\cite{lte},\cite{ciaosr},\cite{clit} primarily utilize LIIF~\cite{liif}. 
Figure~\ref{fig:overview} shows the overview of LIIF-based SR methods~\cite{liif},\cite{lte},\cite{ciaosr},\cite{clit}.
These methods predict an RGB value at a given query pixel position, $\mathbf{x}_q \in \mathbb{R}^2$, from its nearest latent code, $\mathbf{z} \in \mathbb{R}^C$. Note that $\mathbf{x}_q$ denotes a pixel position in the HR image coordinate system.
This latent code, $\mathbf{z}$, is obtained from the latent representation, $E(I^{LR})$, which is extracted from $I^{LR}$ by an image encoder, $E$.
Let $f$ be an implicit function, $I^{HR}$ is formulated as follows:
\begin{equation}
    I^{HR}(\mathbf{x}_q) = f(\mathbf{z}, \mathbf{x}_q - \mathbf{v}^{*}, \mathbf{c}) + I^{LR}_{\uparrow}(\mathbf{x}_q),
\end{equation}
where $\mathbf{v}^* \in \mathbb{R}^2$ is the pixel position of $\mathbf{z}$ in the HR image coordinate system (i.e., not in the coordinate system of the latent representation). The details are shown in Fig.~\ref{fig:coord_system}.
$\mathbf{c} = (\frac{1}{S_x}, \frac{1}{S_y})$ denotes the cell size representing the pixel size ratio between $E(I^{LR})$ and $I^{HR}$. $I^{LR}_{\uparrow}$ denotes the bilinearly-upsampled LR image for residual learning. For simplicity, $\mathbf{c}$ and $I^{LR}_{\uparrow}$ are omitted from the equations as follows:
\begin{equation}
    I^{HR}(\mathbf{x}_q) = f(\mathbf{z}, \mathbf{x}_q - \mathbf{v}^{*})
    \label{eq:liif}
\end{equation}

In the earliest work~\cite{liif}, $f$ is implemented by a Multi-Layer Perceptron (MLP) with ReLU activations, which consists of FC layers, as shown in Fig.~\ref{fig:existings}~(a). 
However, this implementation is biased towards learning low-frequency components, as revealed in~\cite{lte},\cite{siren},\cite{bacon},\cite{spectral}.
To address this issue, Fourier components are employed for redefining Eq.~(\ref{eq:liif}) in ~\cite{lte} as follows:
\begin{equation}
    I^{HR}(\mathbf{x}_q) = g(h(\mathbf{z}, \mathbf{x}_q - \mathbf{v}^*)),
    \label{eq:lte}
\end{equation}
where $h$ denotes a Fourier estimator, providing Fourier components to $g$.
$g$ is a decoding function aggregating the estimated Fourier components to predict an RGB value at $\mathbf{x}_q$.
While several previous methods~\cite{ciaosr},\cite{clit} use Eq.~(\ref{eq:lte}), LTE~\cite{lte}, which is closely related to our proposed method, is briefly presented in what follows.

In LTE, given a local pixel position $\boldsymbol{\delta}$ (i.e., $\boldsymbol{\delta} = \mathbf{x}_q - \mathbf{v}^*$), $h$ is defined as follows:
\begin{eqnarray}
    h(z, \delta) &=& \mathbf{A} \odot
    \begin{bmatrix}
        \rm{cos}(\pi \mathbf{F}\boldsymbol{\delta}) \\
        \rm{sin}(\pi \mathbf{F}\boldsymbol{\delta})
    \end{bmatrix}
    \label{eq:fourier_estimator}
    \\
    \mathbf{A} &=& h_a(\mathbf{z})
    \\
    \mathbf{F} &=& h_f(\mathbf{z})
\end{eqnarray}
where $\mathbf{A} \in \mathbb{R}^{2K}$ and $\mathbf{F} \in \mathbb{R}^{K \times 2}$ denote an amplitude vector and a frequency matrix, respectively. $h_a$ and $h_f$ are an amplitude estimator and a frequency estimator, respectively. $\odot$ denotes an element-wise multiplication operator. $K$ is a hyperparameter that defines the number of Fourier components to be estimated. 
$g$ in Eq.~(\ref{eq:lte}) decodes the estimated $K$ Fourier components into an RGB value at $\mathbf{x}_q$ (i.e., $I^{HR}(\mathbf{x}_q)$).
Each of $h_a$, $h_f$, and $g$ is implemented by an FC layer.

While these estimators and the decoder can be easily implemented by FC layers, as mentioned above, each FC layer consists of a fixed number of input and output channels.
However, such a fixed number of channels can handle only a pre-defined number of Fourier components (i.e., $K$ Fourier components, where $K$ is constant). 

On the other hand, our goal is to control the computational cost and the SR quality by making the number of estimated Fourier components, $K$, variable. Our proposed method utilizes an RNN to allow $K$ to be freely changed at test time.

\subsection{Recurrent Fourier Predictor}

Our proposed implicit function, $f$, is shown in Fig.~\ref{fig:proposed}. 
In our method, for a variable number of Fourier components, an RNN estimates the amplitude vector, $\mathbf{A}_{t} \in \mathbb{R}^2$, and the frequency matrix, $\mathbf{F}_t \in \mathbb{R}^2$, one by one (i.e., a pair of $\mathbf{A}_t$ and $\mathbf{F}_t$ estimated in each recurrence).
The initial hidden state, $\boldsymbol{\alpha}_0$, is derived from the latent code, $\mathbf{z}$, by a mapping function, $\phi$, as illustrated in the upper-left part of Fig.~\ref{fig:proposed}. Let $\psi$ be an RNN unit, our proposed Fourier estimator is rewritten from Eq.~(\ref{eq:fourier_estimator}) to Eq.~(\ref{eq:our_fourier_estimator}) as follows:
\begin{eqnarray}
    \boldsymbol{\alpha}_0 &=& \phi(\mathbf{z})\\
    \boldsymbol{\alpha}_t, \boldsymbol{\beta}_t &=& \psi(\boldsymbol{\alpha}_{t-1}, \gamma(\mathbf{A}_{t-1}, \mathbf{F}_{t-1}))
    \label{eq:our_fourier_estimator} \\
    \mathbf{A}_t &=& h_a(\boldsymbol{\beta}_t)\\
    \mathbf{F}_t &=& h_f(\boldsymbol{\beta}_t)
\end{eqnarray}
where $\gamma$ denotes a function that transforms $\mathbf{A}_{t-1}$ and $\mathbf{F}_{t-1}$ to feature vectors used at $t$-th recurrence. Each of $\phi$ and $\gamma$ is implemented by an FC layer.
This RNN-based Fourier estimator allows us to obtain a variable number of Fourier components by changing the number of recurrences. 

Furthermore, since the estimated Fourier components are fed into the decoding function, it must be capable of taking a variable number of Fourier components.
Instead of a fixed-length FC layer as the decoding function (i.e., $g$ in Eq.~(\ref{eq:lte})), our method leverages the characteristic of the Fourier representation, where the pixel value can be expressed through summation of a variable number of Fourier components. Thus, the RGB value is obtained through a simple summation after the RNN estimates the Fourier components as follows:
\begin{equation}
    I^{HR}(\mathbf{x}_q) = \sum^T_{t=1} \mathbf{A}_t \cdot
    \begin{bmatrix}
        \text{cos}(\pi \mathbf{F}_t\boldsymbol{\delta}) \\
        \text{sin}(\pi \mathbf{F}_t\boldsymbol{\delta})
    \end{bmatrix}
    \label{eq:summation_decoder}
\end{equation}
where $T$ denotes the number of recurrences in the RNN, which is equal to the number of Fourier components used for our RNN-based SR reconstruction.
$\cdot$ denotes the inner product.

\subsection{Training with a variable number of Fourier Components}
\label{sec:random_length_training}

If an RNN is trained only with $T$ recurrences, it is overfitted to $T$ Fourier components and is not generalized to different numbers of Fourier components at test time.
For test-time CQ controllable SR, however, the number of Fourier components is changed at test time, leading to a decline in SR quality in $T'$ Fourier components, where $T' \neq T$, if the RNN is trained only with $T$ recurrences.

For our test-time CQ controllable SR with a variable number of Fourier components, Eq.~(\ref{eq:summation_decoder}) is rewritten to train the RNN with various $T$ as follows:
\begin{equation}
    I^{HR}(\mathbf{x}_q) = \sum^{T\sim \mathcal{U}(1, T_{max})}_{t=1} \mathbf{A}_t \cdot
    \begin{bmatrix}
        \text{cos}(\pi \mathbf{F}_t\boldsymbol{\delta}) \\
        \text{sin}(\pi \mathbf{F}_t\boldsymbol{\delta})
    \end{bmatrix}
    \label{eq:random_length_training}
\end{equation}
where $\mathcal{U}(a, b)$ and $\sim$ denote a uniform discrete distribution over the range from $a$ to $b$ and the random sampling operator, respectively. $T_{max}$ denotes the maximum number of recurrences during training.
In Eq.~(\ref{eq:random_length_training}), an RGB value is predicted with a subset (i.e., $T$ Fourier components) of the $T_{max}$ Fourier components. 
With this RNN training with a variable number of Fourier components, the RNN is expected to achieve high-quality SR reconstruction even with only a subset of $T_{max}$ Fourier components.

With the aforementioned training strategy, it is expected that more dominant Fourier components are estimated in early recurrences. This is because the RNN is trained so that the SR image can be reconstructed better even with fewer Fourier components. Note that it is unknown which Fourier components are estimated in which recurrences; the estimation order fully depends on the training process.

\subsection{Other Details}
\label{sec:others}

\noindent
\textbf{Feature Unfolding:}
To enhance all latent codes in the latent representation (i.e., $E(I^{LR})$), feature unfolding is applied to each latent code, following \cite{liif}. Specifically, $\mathbf{z}_{i,j}$ is concatenated with its 8-neighborhood latent codes as follows:
\begin{equation}
    \hat{\mathbf{z}}_{i,j} = \text{Concat}({\mathbf{z}_{i+l, j+m} | l, m \in \{-1, 0, 1\})},
\end{equation}
where Concat denotes the concatenation of a set of vectors along the channel dimension.
If $(i+l, j+m)$ falls outside $E(I^{LR})$, $\mathbf{z}_{i+l, j+m}$ are padded with zero-vectors.
After this feature unfolding process, $\mathbf{z}_{i,j}$ is replaced with $\hat{\mathbf{z}}_{i,j}$ in all processes in this paper.

\noindent
\textbf{Phase Estimation:}
In SR methods using the Fourier representation, phase information is crucial because it captures the locations of edges and other fine details in an image. In LTE, phase information for each Fourier component is estimated only based on the cell size, $\mathbf{c}$, through an FC layer. The FC layer with a fixed number of channels can be used in LTE because the number of Fourier components is fixed. However, since our proposed method uses a variable number of Fourier components, the phase estimation must also be variable.

For the aforementioned variable number of Fourier components, our method estimates phase information in each recurrence. Let $h_p$ be a phase estimator, Eq.~(\ref{eq:summation_decoder}) is redefined as follows:
\begin{equation}
    I^{HR}(\mathbf{x}_q) = \sum^T_{t=1} \mathbf{A}_t \cdot
    \begin{bmatrix}
        \text{cos}(\pi \mathbf{F}_t\boldsymbol{\delta} + h_p(\mathbf{c}, \mathbf{A}_t, \mathbf{F}_t) \\
        \text{sin}(\pi \mathbf{F}_t\boldsymbol{\delta} + h_p(\mathbf{c}, \mathbf{A}_t, \mathbf{F}_t),
    \end{bmatrix}
    \label{eq:phase_estimator}
\end{equation}
While only $\mathbf{c}$ is fed into the phase estimator in LTE, our phase estimator also takes $\mathbf{A}_t$ and $\mathbf{F}_t$. $\mathbf{A}_t$ and $\mathbf{F}_t$ are not required in LTE because each channel in its FC layer as the phase estimator is implicitly related to each Fourier component (i.e., $\mathbf{A}_t$ and $\mathbf{F}_t$ in Eq.~(\ref{eq:phase_estimator})). This implicit relationship allows us not to feed Fourier components into the phase estimator.
In our RNN-based phase estimator, on the other hand, it is unknown which Fourier component is estimated in each recurrence. To explicitly specify which Fourier component is estimated in each recurrence, $\mathbf{A}_t$ and $\mathbf{F}_t$ are fed into the phase estimator, $h_p$.
In practice, $h_p$ is implemented by an FC layer.

By incorporating the phase estimation process that depends on $\mathbf{A}_t$ and $\mathbf{F}_t$ as described above, our method can correctly use a varying number of Fourier components with phase information.


\section{RESULTS}
\label{sec:results}

\subsection{Implementation Details}

\noindent
{\bf Network architectures:}
The feature extractor, $E$, is given by fine-tuning a pretrained backbone of EDSR~\cite{edsr}. $T_{max}$ is 64, following the previous method~\cite{lte} where the fixed number of Fourier components, $K$, is 64. The RNN unit is implemented as a Linear Transformer~\cite{linear_attention} with four layers. As position encoding for Linear Transformer, our method uses SPE~\cite{spe}, which is a relative position encoding method.

\noindent
{\bf Training:}
At the training time, real-valued scale factors, $S_y$ and $S_x$, are randomly selected, ranging from 1 to 4 in each iteration.
Given an HR image, it is randomly cropped to a $48S_y\times48S_x$ pixel image. This cropped image is then downsampled to $48\times48$ pixels using bicubic interpolation. This downsampled image is regarded as the LR image, which is fed into the SR network. The overall network is trained by randomly selecting 256 query pixel positions in an HR image in each training iteration. The batch size is 16. The model is trained using the Adam optimizer. The learning rate is initialized at 1e-4 and halved every 200 epochs, for a total of 400 epochs. The DIV2K~\cite{div2k} training set is used for network training.

\subsection{Comparison with SoTA Methods}
\label{sec:eval}

\noindent
\textbf{SoTA methods modified with CQ controllability:}
In Sec.~\ref{sec:results}-\ref{sec:eval}, the CQ controllability of our proposed method is evaluated.
Our method is compared with other state-of-the-art arbitrary-scale SR methods, i.e., LTE~\cite{lte}, CiaoSR~\cite{ciaosr}, and CLIT~\cite{clit}, all of which utilize implicit functions. However, since the CQ controllable arbitrary-scale SR is our novel contribution, these state-of-the-art methods do not support CQ control. Therefore, for a fair comparison, these methods are modified to be CQ controllable at test time as follows:
\begin{itemize}
    \item LTE~\cite{lte} estimates Fourier components in the second last layer and then feeds them into the last FC layer for decoding them to an RGB value. Since this last FC layer has a fixed number of channels (i.e., $K$ Fourier components), it prevents LTE from handling a variable number of Fourier components. To use a variable number of Fourier components in our experiments, the last layer of LTE is removed. Instead, a simple sum computation is performed with only some of the Fourier components selected from all the estimated Fourier components.
    Note that the number of the estimated Fourier components in the modified network remains fixed at both training and test time because the FC layer estimates these Fourier components. Therefore, several Fourier components must be selected for CQ control, as mentioned above. This selection process is described in the next paragraph.
    \item CiaoSR~\cite{ciaosr} and CLIT~\cite{clit} do not utilize the Fourier representation. Instead, these methods directly output each RGB value through an FC layer from the latent code.
    To enable CQ control in our experiments, the output FC layer of these methods is replaced with an FC layer that estimates a fixed number of Fourier components. From the estimated Fourier components, a part of them is selected and used to compute an RGB value in the same manner as the modified LTE.
\end{itemize}
These modified networks are trained from scratch, following the training settings prescribed by their papers and original codes.

In the aforementioned modified networks, Fourier components estimated by the last FC layer do not have any inherent order or priority.
While only a subset of the estimated Fourier components can be used for SR reconstruction, inappropriate selection may degrade SR quality.
In our experiments, the following selections are used for the modified networks:
\begin{itemize}
    \item \textbf{Random:} Fourier components are randomly selected from all the estimated Fourier components and used for computing an RGB value. This process is repeated 100 times, and their mean is regarded as the output RGB value.
    \item \textbf{Descending:} Fourier components are selected based on the norm of the two amplitude values (i.e., those of the sine and cosine waves), from largest to smallest. This selection is based on the assumption that Fourier components with larger amplitudes are more dominant and important for estimating RGB values.
\end{itemize}

\begin{table}[t]
\centering
\caption{The average PSNR (dB) on the DIV2K dataset with factor 2. The red score is the \red{best} in each column.}
\begin{tabular}{cc||ccccc}
\hline
\multicolumn{2}{c||}{\multirow{2}{*}{SR methods}}               & \multicolumn{5}{c}{\# of Fourier components at test time ($T$)} \\ \cline{3-7} 
\multicolumn{2}{c||}{}                                                       & 64     & 32     & 16     & 8      & 4     \\ \hline
\multicolumn{1}{c|}{\multirow{2}{*}{LTE~\cite{lte}}}        & Random        & \red{34.56}  & 22.88  & 21.09  & 22.55  & 24.24 \\
\multicolumn{1}{c|}{}                                       & Descend    & \red{34.56}  & 22.90  & 20.88  & 20.42  & 21.68 \\ \hline
\multicolumn{1}{c|}{\multirow{2}{*}{CiaoSR~\cite{ciaosr}}}  & Random        & 33.19  & 22.16  & 22.51  & 19.08  & 20.21 \\
\multicolumn{1}{c|}{}                                       & Descend    & 33.19  & 20.87  & 21.87  & 20.51  & 21.23 \\ \hline
\multicolumn{1}{c|}{\multirow{2}{*}{CLIT~\cite{clit}}}      & Random        & 32.95  & 21.69  & 19.92  & 20.61  & 22.52 \\
\multicolumn{1}{c|}{}                                       & Descend    & 32.95  & 20.12  & 19.18  & 21.56  & 21.84 \\ \hline
\multicolumn{2}{c||}{Ours}                                                   & 34.14  & \red{34.01}  & \red{33.51}  & \red{33.00}  & \red{32.16} \\ \hline
\end{tabular}
\label{tab:div_x2}
\medskip
\centering
\caption{The average PSNR (dB) on the DIV2K dataset with factor 3. The red score is the \red{best} in each column.}
\begin{tabular}{cc||ccccc}
\hline
\multicolumn{2}{c||}{\multirow{2}{*}{SR methods}}               & \multicolumn{5}{c}{\# of Fourier components at test time ($T$)} \\ \cline{3-7} 
\multicolumn{2}{c||}{}                                     & 64     & 32     & 16     & 8      & 4     \\ \hline
\multicolumn{1}{c|}{\multirow{2}{*}{LTE~\cite{lte}}}    & Random     & \red{30.78}  & 23.38  & 20.59  & 22.03  & 18.72 \\
\multicolumn{1}{c|}{}                        & Descend & \red{30.78}  & 23.57  & 20.74  & 19.02  & 18.99 \\ \hline
\multicolumn{1}{c|}{\multirow{2}{*}{CiaoSR~\cite{ciaosr}}} & Random     & 30.14  & 23.05  & 20.81  & 20.14  & 22.58 \\
\multicolumn{1}{c|}{}                        & Descend & 30.14  & 22.11  & 21.12  & 20.51  & 20.19 \\ \hline
\multicolumn{1}{c|}{\multirow{2}{*}{CLIT~\cite{clit}}}   & Random     & 29.78  & 20.86  & 21.10  & 20.91  & 21.14 \\
\multicolumn{1}{c|}{}                        & Descend & 29.78  & 21.91  & 20.32  & 21.06  & 20.89 \\ \hline
\multicolumn{2}{c||}{Ours}                                 & 30.51  & \red{30.36}  & \red{30.30}  & \red{29.41}  & \red{27.56} \\ \hline
\end{tabular}
\label{tab:div_x3}
\medskip
\centering
\caption{The average PSNR (dB) on the DIV2K dataset with factor 4. The red score is the \red{best} in each column.}
\begin{tabular}{cc||ccccc}
\hline
\multicolumn{2}{c||}{\multirow{2}{*}{SR methods}}               & \multicolumn{5}{c}{\# of Fourier components at test time ($T$)} \\ \cline{3-7} 
\multicolumn{2}{c||}{}                                     & 64     & 32     & 16     & 8      & 4     \\ \hline
\multicolumn{1}{c|}{\multirow{2}{*}{LTE~\cite{lte}}}    & Random     & \red{29.22}  & 21.56  & 18.74  & 19.82  & 21.43 \\
\multicolumn{1}{c|}{}                        & Descend & \red{29.22}  & 22.28  & 18.86  & 16.22  & 17.01 \\ \hline
\multicolumn{1}{c|}{\multirow{2}{*}{CiaoSR~\cite{ciaosr}}} & Random     & 28.15  & 20.01  & 18.50  & 18.21  & 21.55 \\
\multicolumn{1}{c|}{}                        & Descend & 28.15  & 22.23  & 20.54  & 19.42  & 18.30 \\ \hline
\multicolumn{1}{c|}{\multirow{2}{*}{CLIT~\cite{clit}}}   & Random     & 27.91  & 18.90  & 19.21  & 19.81  & 20.99 \\
\multicolumn{1}{c|}{}                        & Descend & 27.91  & 20.61  & 19.15  & 20.15  & 19.51 \\ \hline
\multicolumn{2}{c||}{Ours}                                 & 28.61  & \red{28.51}  & \red{28.04}  & \red{26.49}  & \red{25.81} \\ \hline
\end{tabular}
\label{tab:div_x4}
\end{table}

\begin{table}[t]
\centering
\caption{The average PSNR (dB) on the BSD100 dataset with factor 4. The red score is the \red{best} in each column.}
\begin{tabular}{cc||ccccc}
\hline
\multicolumn{2}{c||}{\multirow{2}{*}{SR methods}}               & \multicolumn{5}{c}{\# of Fourier components at test time ($T$)} \\ \cline{3-7} 
\multicolumn{2}{c||}{}                                     & 64     & 32     & 16     & 8      & 4     \\ \hline
\multicolumn{1}{c|}{\multirow{2}{*}{LTE}}    & Random     & \red{26.92}  & 20.55  & 17.91  & 18.12  & 20.23 \\
\multicolumn{1}{c|}{}                        & Descend & \red{26.92}  & 21.26  & 19.46  & 17.61  & 18.07 \\ \hline
\multicolumn{1}{c|}{\multirow{2}{*}{CiaoSR}} & Random     & 26.25  & 19.71  & 17.58  & 17.82  & 19.72 \\
\multicolumn{1}{c|}{}                        & Descend & 26.25  & 21.16  & 19.77  & 18.53  & 18.53 \\ \hline
\multicolumn{1}{c|}{\multirow{2}{*}{CLIT}}   & Random     & 25.96  & 20.55  & 18.46  & 20.30  & 20.34 \\
\multicolumn{1}{c|}{}                        & Descend & 25.96  & 22.17  & 20.15  & 19.61  & 19.59 \\ \hline
\multicolumn{2}{c||}{Ours}                                 & 26.62  & \red{26.58}  & \red{26.19}  & \red{25.20}  & \red{24.71} \\ \hline
\end{tabular}
\label{tab:bsd_x4}
\medskip
\centering
\caption{Average PSNR (dB) on the Urban100 dataset with factor 4. The red score is the \red{best} in each column.}
\begin{tabular}{cc||ccccc}
\hline
\multicolumn{2}{c||}{\multirow{2}{*}{SR methods}}               & \multicolumn{5}{c}{\# of Fourier components at test time ($T$)} \\ \cline{3-7} 
\multicolumn{2}{c||}{}                                     & 64     & 32     & 16     & 8      & 4     \\ \hline
\multicolumn{1}{c|}{\multirow{2}{*}{LTE~\cite{lte}}}    & Random     & \red{26.80}  & 20.91  & 18.07  & 19.21  & 20.43 \\
\multicolumn{1}{c|}{}                        & Descend & \red{26.80}  & 20.99  & 20.16  & 19.86  & 17.90 \\ \hline
\multicolumn{1}{c|}{\multirow{2}{*}{CiaoSR~\cite{ciaosr}}} & Random     & 25.85  & 21.06  & 21.96  & 19.51  & 20.83 \\
\multicolumn{1}{c|}{}                        & Descend & 25.85  & 22.22  & 20.61  & 19.10  & 20.11 \\ \hline
\multicolumn{1}{c|}{\multirow{2}{*}{CLIT~\cite{clit}}}   & Random     & 25.68  & 19.91  & 19.05  & 19.79  & 21.51 \\
\multicolumn{1}{c|}{}                        & Descend & 25.68  & 20.05  & 19.52  & 19.63  & 20.17 \\ \hline
\multicolumn{2}{c||}{Ours}                                 & 26.52  & \red{26.47}  & \red{25.80}  & \red{25.12}  & \red{24.90} \\ \hline
\end{tabular}
\label{tab:urban_x4}
\end{table}

\noindent
\textbf{Quantitative results:}
Tables~\ref{tab:div_x2}, \ref{tab:div_x3}, and \ref{tab:div_x4} show the quantitative results on the DIV2K~\cite{div2k} validation set with scale factors 2, 3, and 4, respectively. As shown in these tables, if the number of Fourier components used at test time, $T$, is equal to their maximum number used at training time (i.e., $T_{max} = 64$), LTE is the best in all scale factors. This result may be attributed to the model complexity of our proposed method, which uses an RNN instead of the simpler FC layer used in LTE. This model complexity may decrease performance.
However, the PSNR gaps between LTE and our method are not significant (i.e., 0.42, 0.27, and 0.61 in scale factors 2, 3, and 4, respectively).

On the other hand, we can also see that our proposed method maintains a high PSNR score even with fewer Fourier components (i.e., $T \in \{ 32, 16, 8, 4 \}$), while the PSNR scores of the other methods significantly decrease as the number of Fourier components is reduced.
For example, in scale factor 4, the PSNR gaps between LTE and our method are 6.95, 9.30, 6.67, and 4.38 with $T=$ 32, 16, 8, and 4, respectively. These PSNR gaps are significantly larger than those with $T=64$, in which LTE is better than our method as shown in the last paragraph.

Additionally, in our method, the PSNR score increases monotonically as the number of Fourier components increases. This property allows us to intuitively control CQ.

The advantages of our method demonstrated above are acquired by our RNN-based CQ control, which cannot be achieved by FC layer-based SR networks used in the other SoTA arbitrary-scale SR methods.

To further validate the CQ controllability of our proposed method, we experimented on two additional datasets: BSD100~\cite{bsd} and Urban109~\cite{urban}. The quantitative results on these datasets are shown in Tables \ref{tab:bsd_x4} and \ref{tab:urban_x4}. From these tables, we can see that our method is robust to a decline in Fourier components, similar to the results of the DIV2K dataset. These results demonstrate good CQ controllability of our proposed method.

\begin{figure*}
    \centering
    \includegraphics[width=\linewidth]{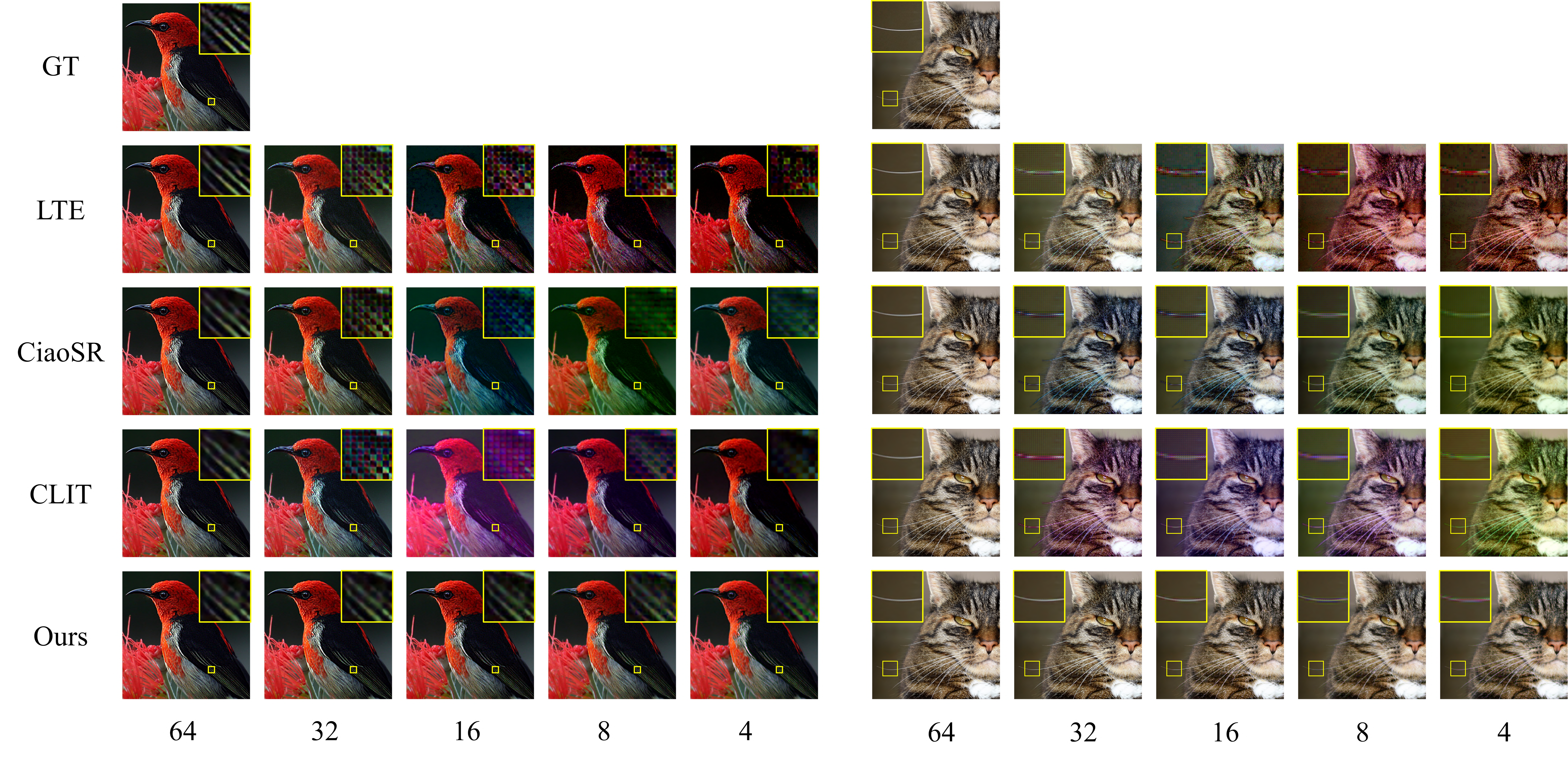}
    \caption{Visual comparisons on DIV2K with factor 4. The yellow rectangular areas are enlarged for better visualization.}
    \label{fig:qualitative}
\end{figure*}

\noindent
\textbf{Visual results:}
Visual results are shown in Fig.~\ref{fig:qualitative}. From Fig.~\ref{fig:qualitative}, we can see that color shifts and severe artifacts occur in the previous methods~\cite{lte, ciaosr, clit} when the number of Fourier components is reduced. This is because these methods are designed with a fixed number of Fourier components at training time due to the FC layer. Consequently, each Fourier component strongly depends on other Fourier components within the FC layer. As a result, reducing the number of Fourier components disrupts this dependency and leads to color shifts and artifacts.

In contrast, our proposed method maintains reasonable SR quality even with fewer Fourier components. This proves that our proposed architecture and training strategy can generalize well even with fewer Fourier components and highlights the CQ controllability of our proposed method.

\begin{figure*}
    \centering
    \includegraphics[width=\linewidth]{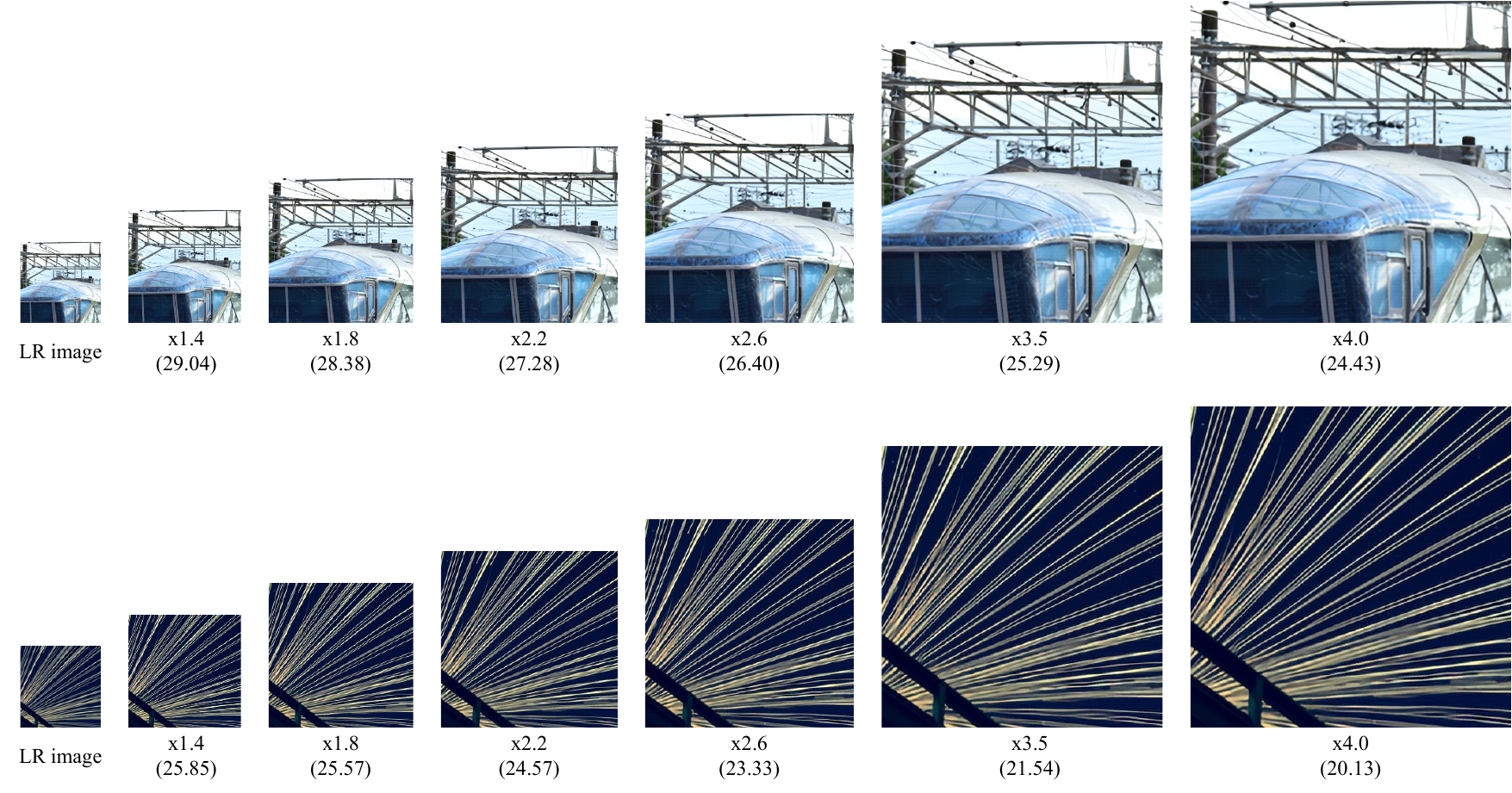}
    \caption{Visual results of our SR method with non-integer scale factors. The input images are obtained by downsampling HR images (i.e., DIV2K image ID=820, 828) to 1/4 of their original size using bicubic interpolation. PSNR values for each SR image are shown in parentheses. The PSNR score is calculated by comparing the SR images with pseudo GTs, obtained by downsampling the original HR images to the corresponding size.}
    \label{fig:non_integer}
\end{figure*}

\subsection{Analysis of Proposed Method}

\noindent
\textbf{Visual results with non-integer scaling factors:} The visual results with non-integer scale factors of our method are shown in Fig~\ref{fig:non_integer}. The results shows that our method achieves consistent reconstruction across different scale factors. This demonstrate that our method is capable of arbitrary scaling factor and CQ controllability simultaniously.

\noindent
\textbf{Effectiveness of training with a variable number of Fourier components:}
To validate the effect of training with a variable number of Fourier components described in Sec.~\ref{sec:methods}-\ref{sec:random_length_training}, we compare the PSNR scores between our method with and without this variable-length training. Without variable-length training, all Fourier components estimated by the RNN are summed up to obtain RGB values during training. Specifically, $T$ in Eq.~(\ref{eq:random_length_training}) is $T_{max} (=64)$ in all training iterations.

\begin{figure}[t]
    \centering
    \includegraphics[width=\linewidth]{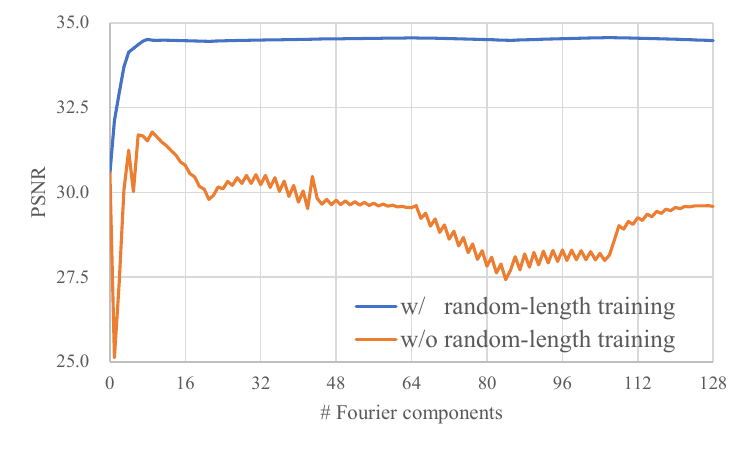}
    \caption{Comparison of our models trained with and without variable-length training.
    In this graph, PSNR scores are plotted by varying the number of Fourier components used at test time from 1 to 128. For these results, only the first five images in the DIV2K validation set (i.e., image IDs=801, 802, 803, 804, 805) are used.}
    \label{fig:num_fourier}
\end{figure}

The results are shown in Fig.~\ref{fig:num_fourier}. Without variable-length training, the PSNR score changes unstably as the number of recurrences changes. This is because the SR model is not generalized well across different numbers of recurrences at training time. Furthermore, even when the number of recurrences is $T_{max} = 64$, which is the number of Fourier components used at the training time, PSNR is still lower compared to the result with variable-length training. 

\noindent
\textbf{Training with different $T_{max}$:}
As described in Sec.~\ref{sec:methods}-\ref{sec:random_length_training}, our proposed method is trained by using only a subset of $T_{max}$ Fourier components to control CQ. Since $T_{max}$ is a hyperparameter, we investigate how $T_{max}$ affects the SR quality. For this evaluation, our model is trained independently with different $T_{max}$ values (i.e., $T_{max} \in \{ 4, 8, 16, 32, 64\}$). With each trained model, SR images are reconstructed with different numbers of Fourier components at test time. The experimental results are shown in Table~\ref{tab:T_max}.

This table shows that the highest PSNR score is obtained when the number of Fourier components at test time is equal to $T_{max}$.
This result reveals the potential for further improvement in CQ controllability depending on the upper limitation of the condition. For example, if the computational resource of a user is limited so that $T=8$ Fourier components can be used at most at test time, the SR model trained with $T_{max}=8$ is the best choice, while any $T$ Fourier components (where $T \leq T_{max}$) can be used for CQ control with this SR model.

Moreover, PSNR degrades when the number of Fourier components exceeds $T_{max}$. This degradation is natural since the model is not trained beyond $T_{max}$.
However, since it may be possible to suppress the performance degradation occurred by recurrences beyond the training range in RNNs as proposed in~\cite{extrapolate1,extrapolate2}, this is one of the challenging future research directions.

\begin{table}[t]
\centering
\caption{Average PSNR scores of different $T_{max}$. The red score is the \red{best} in each column.}
\begin{tabular}{cc||ccccc}
\hline
\multicolumn{2}{c||}{\multirow{2}{*}{}}             & \multicolumn{5}{c}{\# of Fourier components at test time ($T$)} \\ \cline{3-7} 
\multicolumn{2}{c||}{}                                   & 64     & 32     & 16     & 8      & 4     \\ \hline
\multirow{5}{*}{$T_{max}$}                                 & 64 & \red{28.62} & 28.54 & 28.43 & 27.86 & 26.18 \\
                                                        & 32 & 28.53 & \red{28.59} & 28.26 & 27.48 & 25.99 \\
                                                        & 16 & 28.31 & 28.44 & \red{28.59} & 28.12 & 26.33 \\
                                                        & 8  & 27.97 & 28.37 & 28.41 & \red{28.46} & 27.98 \\
                                                        & 4  & 19.81 & 25.01 & 27.55 & 27.80 & \red{28.40} \\ \hline
\end{tabular}
\label{tab:T_max}
\end{table}

\noindent
\textbf{Position Encoding:}
The attention mechanism employed by Transformers does not inherently consider the positional information of input data. Consequently, for tasks where positional information is crucial (e.g., natural language processing and image classification), the usefulness of position encoding has been extensively validated~\cite{pos_emb1},\cite{pos_emb2}.
However, the efficacy of position encoding in the Fourier domain remains unexplored. We experiment with the three different position encoding methods, as shown in Table~\ref{tab:pos_emb}.

\begin{table}[t]
\centering
\caption{Average PSNR (dB) of different position encoding. The red value indicates the \red{best} score in each column.}
\begin{tabular}{cc||ccccc}
\hline
\multicolumn{2}{c||}{\multirow{2}{*}{Position Encoding}}             & \multicolumn{5}{c}{\# of Fourier components at test time ($T$)} \\ \cline{3-7} 
\multicolumn{2}{c||}{Methods}                                   & 64     & 32     & 16     & 8      & 4     \\ \hline
\multicolumn{2}{c||}{None}              & 28.61 & 28.51 & 28.04 & 26.49 & 25.81 \\ \hline
\multirow{2}{*}{Absolute} & Sinusoidal & 28.62 & 28.51 & 28.10 & 26.57 & 25.84 \\
                          & Learned    & \red{28.63} & 27.00 & 26.86 & 26.27 & 24.51 \\ \hline
\multicolumn{2}{c||}{Relative}          & 28.62 & \red{28.54} & \red{28.43} & \red{27.86} & \red{26.18} \\ \hline
\end{tabular}
\label{tab:pos_emb}
\end{table}

This table shows that absolute position encoding provides no improvement compared to our method without any position encoding. In contrast, relative position encoding improves the PSNR score with fewer Fourier components (i.e., 32, 16, 8, and 4 Fourier components). 
This is because of the nature of each position encoding method. Absolute position encoding assigns a unique embedding to each position independently. This independence means that training for each position does not generalize to others. On the other hand, relative position encoding considers the relative positions so that the relationships trained in one set of Fourier components influence others. For instance, the knowledge used for predicting attention between the 1st and 2nd Fourier components can also be applied to predicting attention between the 31st and 32nd components. This shared knowledge contributes the performance improvement with fewer Fourier components.
These results reveal that the generalized knowledge of Fourier components can be represented by relative position encoding.


\section{CONCLUSION}

In this paper, we proposed a novel method for arbitrary-scale super-resolution (SR) that leverages Recurrent Neural Networks (RNNs) to estimate Fourier components, enabling test time controllability between cost and quality. Quantitative evaluations demonstrate that our method retains high PSNR even with fewer Fourier components, highlighting its cost-and-quality controllability. Additionally, our qualitative analysis shows that, unlike previous methods, our method produces consistent and artifact-free results even with fewer Fourier components.
Future work includes how to generalize the SR model to more Fourier components than the number of max RNN recurrences used at training time. Additionally, we plan to explore autoregressive models~\cite{} instead of RNNs to enhance generative capabilities and further improve the performance of our method.


\bibliographystyle{ieeetr}
\bibliography{reference}

\begin{IEEEbiography}[{\includegraphics[width=1in,height=1.25in,clip,keepaspectratio]{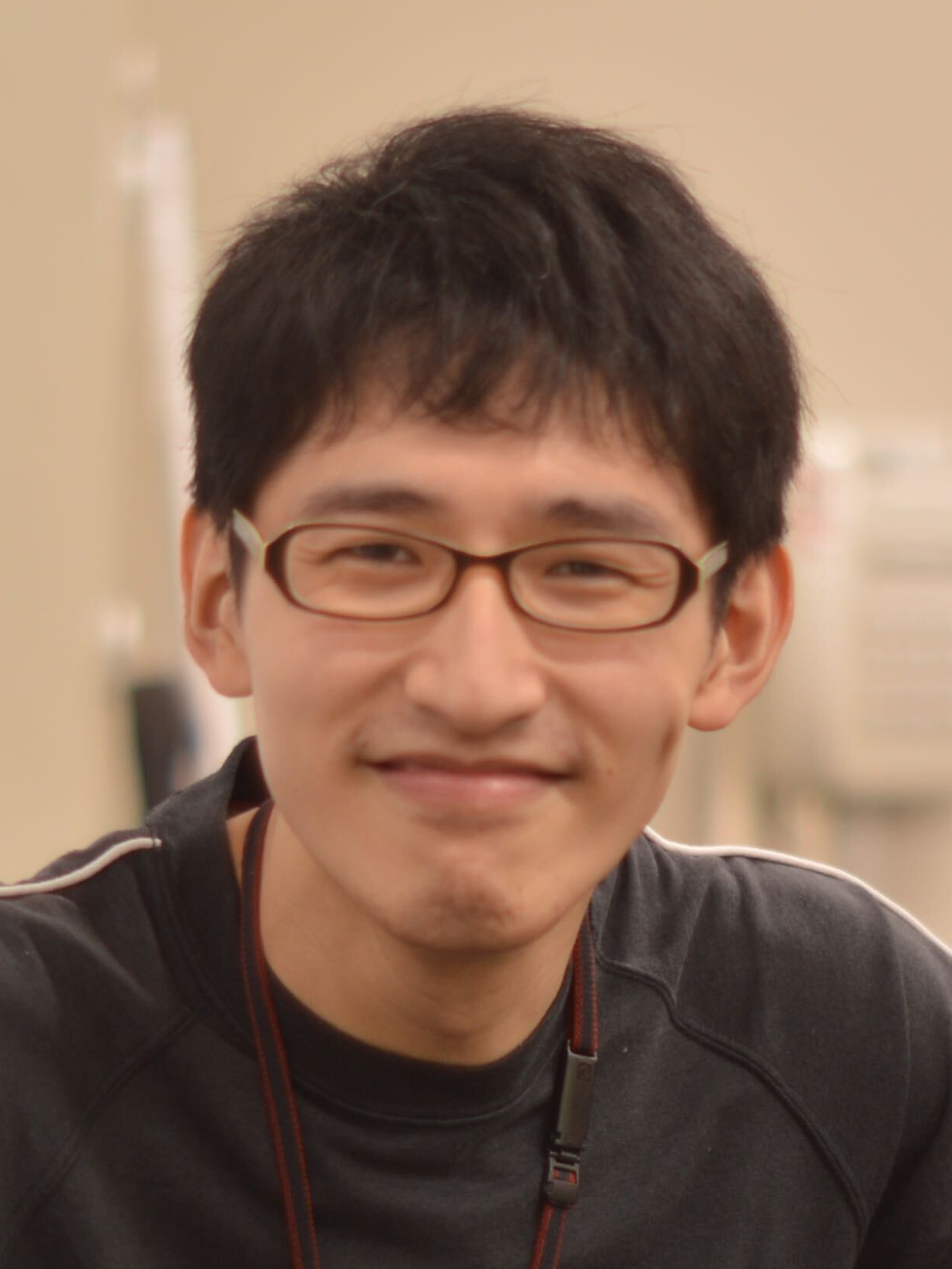}}]
{KAZUTOSHI AKITA}~received the B.E degree in 2019, and the M.E. degree in 2021 from Toyota Technological Institute, Japan (TTI-J), and he is currently pursuing the Ph.D. degree at TTI-J. His research interests include object detection, super-resolution, and their coupling.
\end{IEEEbiography}

\begin{IEEEbiography}[{\includegraphics[width=1in,height=1.25in,clip,keepaspectratio]{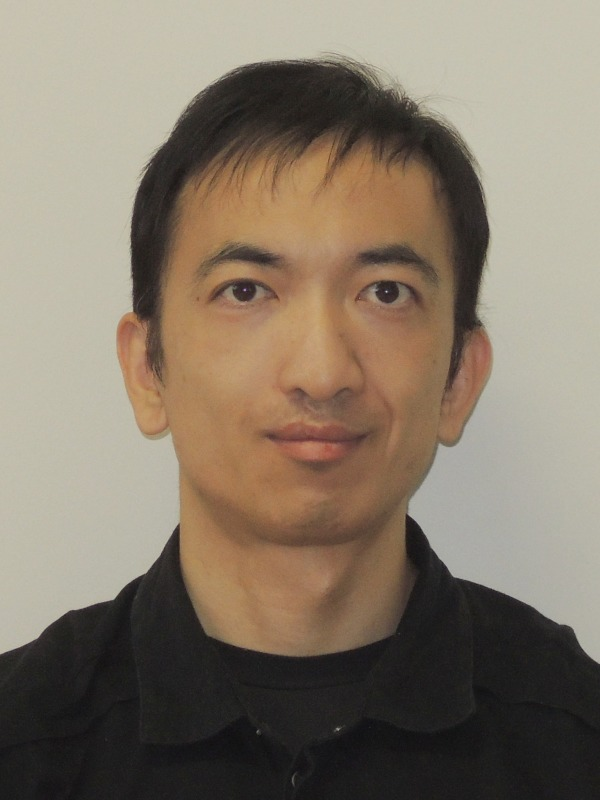}}]
{NORIMICHI UKITA}~received the B.E. and M.E. degrees in information engineering from Okayama University, Japan, in 1996 and 1998, respectively, and the Ph.D. degree in Informatics from Kyoto University,
Japan, in 2001. From 2001 to 2016, he was an assistant professor (2001 to 2007) and an associate
professor (2007-2016) with the graduate school of information science, Nara Institute of Science and
Technology, Japan. In 2016, he became a professor at Toyota Technological Institute, Japan. He was a research scientist of Precursory Research for Embryonic Science and Technology, Japan Science and Technology Agency, during 2002–2006, and a visiting research scientist at Carnegie Mellon University during 2007–2009. Currently, he is also an adjunct professor at Toyota Technological Institute at Chicago. Prof. Ukita’s awards include the excellent paper award of IEICE (1999), the winner award in NTIRE 2018 challenge on image super-resolution, 1st place in PIRM 2018 perceptual SR challenge, the best poster award in MVA2019, and the best practical paper award in MVA2021.
\end{IEEEbiography}

\vfill\pagebreak

\end{document}